\documentclass[final,letterpaper,twoside,12pt]{article}
\pdfoutput=1

\usepackage{color}
\usepackage{graphicx}
\usepackage{graphics}
\usepackage{epsfig}

\newcommand{\be}{\begin{eqnarray}}
\newcommand{\ee}{\end{eqnarray}}
\newcommand{\wt}[1]{\widetilde{#1}}
\newcommand{\mt}{m_{3/2}}
\newcommand{\lt}{\lambda_T}
\newcommand{\ld}{\lambda_D}
\newcommand{\dm}{\Delta M}
\newcommand{\mq}{m_{\tilde{q}}}
\newcommand{\ml}{m_{\tilde{l}}}
\newcommand{\bu}{B \mu}

\newcommand{\zu}{z_u}
\newcommand{\zd}{z_d}

\newcommand{\mhu}{m^2_{h_u}}
\newcommand{\mhd}{m^2_{h_d}}
\newcommand{\jhu}{\emph{Dept.~of Physics \& Astronomy, Johns Hopkins University, Baltimore, MD  21218, USA}}

\begin{document}
\vspace{-5cm}
\title{Gaugomaly Mediation Revisited}
\author{Arpit Gupta\thanks{arpit@pha.jhu.edu}, David E. Kaplan\thanks{dkaplan@pha.jhu.edu}, Tom Zorawski\thanks{tz137@pha.jhu.edu}\\ \\ \small{\jhu}}

\date{}

\maketitle

\begin{abstract}
\hspace{0cm}
Most generic models of hidden sector supersymmetry breaking do not feature singlets, and gauginos obtain masses from anomaly mediated supersymmetry breaking. If one desires a natural model, then the dominant contribution to scalar masses should be of the same order, i.e. also from AMSB. However, pure AMSB models suffer from the tachyonic slepton problem. Moreover, there is a large splitting between the gluino and the wino LSP masses resulting in tight exclusion limits from typical superpartner searches. We introduce messenger fields into this framework to obtain a hybrid theory of gauge and anomaly mediation, solving both problems simultaneously.  Specifically, we find any number of vector-like messenger fields (allowed by GUT unification) compress the predicted gaugino spectrum when their masses come from the Giudice-Masiero mechanism.  This more compressed spectrum is less constrained by LHC searches and allows for lighter gluinos.
In addition to the model, we present gaugino pole mass equations that differ from (and correct) the original literature.
\end{abstract}

\section{Introduction}
Weak-scale supersymmetry (SUSY) is an attractive framework for physics beyond the standard model (SM) as it provides a natural solution to the hierarchy problem as well as a dark matter candidate. However, it faces several challenges. To explain the dearth of SUSY signals at past and present colliders, SUSY has to be broken to give superpartner masses that evade current bounds. A successful SUSY breaking mechanism has to pass constraints related to flavor-changing neutral currents (FCNC) and should preserve gauge coupling unification. Two SUSY breaking mechanisms that satisfy these requirements are anomaly-mediated supersymmetry breaking (AMSB) and gauge-mediated supersymmetry breaking (GMSB), each with its own set of accompanying problems. \\
\indent AMSB is a very appealing SUSY breaking mechanism since it is flavor blind, UV insensitive, and highly predictive~\cite{amsb1,amsb2}. In the minimal AMSB setup, the ratio of the gluino mass to the LSP wino mass is about a factor of ten, meaning that this model is well covered by conventional SUSY searches that rely on hard jets and large missing transverse energy. The null results from such searches based on analyses of all of the approximately 5 $fb^{-1}$ of data~\cite{cms,atlas,atf} collected at the LHC in 2011 and some of the 2012 data~\cite{cms12,at12_6,at12_13} have placed a lower bound on the gluino mass of about 1.4 TeV for many vanilla models, leading to increased tension with naturalness. Furthermore, AMSB models suffer from the tachyonic slepton problem and the $\mu$ problem. \\
\indent GMSB is another widely studied SUSY breaking mechanism. SUSY breaking is communicated from a hidden sector to the superpartner fields via messenger fields that have interactions with SM gauge bosons. To preserve gauge coupling unification, the messengers should come in complete representations of $SU(5)$. GMSB models are predictive and do not suffer from flavor problems. However, they have a gravitino LSP which can be a problem in the context of cosmology. In addition, they also suffer from the $\mu$ problem. \\
\indent  We believe, however, that the most realistic models can be obtained by combining AMSB with GMSB. This approach is not new. Refs.~\cite{chacko, nws} first showed that the D-type gauge mediation of Poppitz-Trivedi~\cite{pop} could be simply combined with AMSB to solve the tachyonic slepton problem, although they did not specify the origin of the messenger masses. Ref.~\cite{ga} studied this further and gave it the name  `gaugomaly' mediation. On a different front, with the goal of solving the tachyonic slepton problem,  ref.~\cite{nw} developed extended anomaly mediation (EAM) by arranging for the messengers to get masses directly from anomaly mediation through Giudice-Masiero type (GM) terms~\cite{gm}. This deflects the gaugino masses off of the AMSB trajectory, indirectly changing the scalar masses through running. The EAM setup was itself extended in~\cite{luty1, luty2} with the addition of a singlet to yield realistic spectra. \\
\indent We take the approach that such singlets are unnatural, so we are led to consider EAM with the D-type GMSB of gaugomaly mediation, which surprisingly has not been explored previously. In addition, we investigate the effect of the messengers in compressing the gaugino spectrum, an interesting aspect not discussed in the references above that deserves attention on its own. Integrating out the messengers takes the gaugino masses off the AMSB trajectory and gives threshold corrections that modify the masses at leading order, with the ratios sensitive to the number of messenger pairs. The result is a compressed gaugino spectrum with the mass splitting between the gluino and the LSP depending on the number of messenger pairs. The limits on the allowed gaugino masses are significantly weakened due to the squeezing of the spectrum. For example, an update of the compressed SUSY analysis of~\cite{lec} based on the full 2011 dataset shows that for splittings of about 50 GeV,  gluino masses as small as 550 GeV are still allowed~\cite{atf} (this still holds when the 2012 data is considered; for the best limit, see~\cite{at12_6}). This framework can also bring models that are otherwise beyond the reach of the LHC to within its reach in certain cases as the gluino becomes only 1-2 times heavier than the LSP (the wino or the bino depending  on the number of messengers). \\
\indent This paper is structured as follows. First, we briefly review AMSB. Then, in section 3, we review the calculation of the gaugino pole masses in the minimal AMSB framework. Although this has already been discussed in~\cite{gg}, we present the details here because our equations differ slightly.  We introduce messengers in section 4 in the context of D-type gauge mediation as a solution to the tachyonic slepton problem. In section 5, we discuss the gaugino spectra and related phenomenology independently from the scalars, focusing on the compression. Then, in section 6, assuming that the $\mu$ problem has been solved, we give complete example spectra with the scalars. We conclude in section 7 by presenting a simple (and rough) solution to the $\mu$ problem in this framework and briefly discuss how it can be improved by incorporating some ideas found in the literature.

\section{Review of AMSB}

In the pure AMSB case, the gaugino masses are generated at the one loop level and the scalar masses squared are generated at two loops. The equations for these soft masses are the solutions to the RGEs for the gauginos and sfermions and thus are valid at all energy scales. The running gaugino masses are given by 

\begin{equation}\label{AMSB_1}
m_{i}=\frac{\beta (g_{i})}{g_{i}} m_{3/2}
\end{equation}

\noindent This formula is valid at all scales, meaning that if we wish to calculate the masses in the IR we only need to know the values of the couplings at that scale.  Using the one-loop beta functions, we obtain

\be
m_{1} &=&  11 \frac{\alpha_1}{4 \pi} m_{3/2}  \label{bino}\\
m_{2} &=& \frac{\alpha_2}{4 \pi} m_{3/2}  \label{wino}\\
m_{3 } &=&  -3 \frac{\alpha_3}{4 \pi} m_{3/2}.  \label{gluino}
\ee

\noindent Note that throughout this paper we use the non-GUT normalization for hypercharge. Plugging in the weak scale values of the $\alpha_i$,  we find that the mass ratios $m_1:m_2:|m_3|$ are approximately $3.3:1:10$. We assume here that $\mu$ is larger than any of the gaugino masses, so that the wino is the LSP in this minimal case. This large splitting between the gluino and the wino means that energetic jets will be produced in the cascade decay (we discuss phenomenology in a later section), leading to tight constraints on the masses. However, as pointed out in~\cite{gg}, quantum corrections here are significant, so we consider the pole masses computed at NLO. We discuss the calculation of the pole masses for the gauginos in the subsequent section. \\ 
\indent The scalar masses in AMSB are given by 
\begin{equation}\label{AMSB_1}
m^2_i = -\frac{1}{4}\left(\frac{d \gamma_i}{dg_j} \beta_{g_j} + \frac{d \gamma_i}{dy_j} \beta_{y_j} \right)m_{3/2}^2,
\end{equation}

\noindent where $\gamma$ is the corresponding anomalous dimension, and $\beta_g$, $\beta_y$ are the gauge coupling and yukawa coupling beta functions, respectively. Since the sleptons are charged only under non-asymptotically free gauge groups, they will have negative squared masses. This is the well-known tachyonic slepton problem in AMSB. 

\section{Gaugino Pole Masses}
The full NLO expression includes contributions coming from $\alpha_3$ and $y_t$ in the two-loop beta functions as well as self-energy corrections. For the one-loop self-energies, our analysis follows closely the steps presented in~\cite{bagger}.

\subsection{Gauge Loops}
The contribution to the self-energy from gauge boson loops is given by

\be
\left(\frac{\Delta m_i}{m_i}\right)_{gauge} = \frac{\alpha_i}{4 \pi} C(G_i) \left[4 B_0(m_i,m_{\psi},m_{\phi})-2B_1(m_i,m_{\psi},m_{\phi})\right] , 
\ee

\noindent where the Veltman-Passarino functions (defined as in~\cite{bagger}) are 

\be
B_0(m_i,m_{\psi},m_{\phi}) &=& - \int_{0}^{1} dx \ln \frac{(1-x) m_{\psi}^2+x m_{\phi}^2-x(1-x) m_i^2}{Q^2}\\
B_1(m_i,m_{\psi},m_{\phi}) &=& - \int_{0}^{1} dx \, x \ln \frac{(1-x) m_{\psi}^2+x m_{\phi}^2-x(1-x) m_i^2}{Q^2},
\ee

\noindent with $m_{\psi}$ the mass of the fermion in the loop, $Q$ is the renormalization scale, and $m_{\phi}$ is the mass of the boson in the loop (vector or scalar). Throughout most of this paper, except where noted otherwise, we work in the $\overline{DR}$ scheme. For the gluino, $m_{\phi} = 0$ (the bosons in the loop are massless gluons), and we can evaluate the integrals directly, yielding 

\be
\left(\frac{\Delta m_3}{m_3}\right)_{gauge} = \frac{\alpha_3}{4 \pi} C(G_3) \left(5+3 \ln \frac{Q^2}{m_3^2}\right).
\ee

For the wino, we must use the full $B$ functions, since $m_W$ is of order $m_2$. The amount of diagrams differs for the neutral and charged wino (in fact, this will be the main source of the splitting between the lightest chargino and the LSP as described later). Here we use the empirical fit of~\cite{bagger} for the neutralino result. 

\subsection{Matter Loops}
Next, we consider the contributions of the matter content to the gaugino masses. We assume that all squarks are degenerate with mass $\mq$, and that all sleptons are degenerate with mass $\ml$, with $ \mq, \ml > m_3$ (this will be important when discussing the phenomenology). In addition, we approximate the fermion masses as zero. For the gluino, only quark/squark loops contribute, giving

\be
\left(\frac{\Delta m_3}{m_3}\right)_{matter} &=& -12 \frac{\alpha_3}{4 \pi} B_1(m_3,0,\mq),\\
- B_1(m_3,0,\mq) &=& -\frac{1}{2} \ln \frac{Q^2}{m_3^2} +I,\\
I &=&  \int_{0}^{1} dx \, x \ln [r x-x(1-x)], \quad r = \left(\frac{\mq}{m_3}\right)^2.
\ee

It is straightforward to evaluate $I$

\be
I = \frac{1}{2} \left(-2+r+(r-1)^2 \ln |1-r| - (r-2) r \ln r \right).
\ee

\noindent Since the top mass is fairly large, we should examine whether neglecting it is a good approximation. Including the top mass gives an extra contribution $\frac{\alpha_3}{4 \pi} \frac{m_t}{m_3}f(m_{\tilde{t}_1},m_{\tilde{t}_2})$, where $f$ is a function of the stop masses that is $\sim 1$. Since in our case $m_3 \sim 1$ TeV, this term is indeed suppressed. \\
\indent For the wino and bino, we divide the matter contributions into fermion/sfermion, chargino/charged Higgs, and neutralino/neutral Higgs pieces. The sfermion piece is

\be
\left(\frac{\Delta m_i}{m_i}\right)_{sfermion} &=& -2 \frac{\alpha_i}{4 \pi} S(R_{\Phi})B_1(m_i,0,m_\phi),
\ee
summing over all chiral supermultiplets $\Phi = (\phi,\psi)$ that couple to the wino/bino. To simplify $B_1$, we can take the wino/bino mass to be approximately zero. At first sight this doesn't seem to be valid, since the gaugino and sfermion masses are both of roughly the same order. However, since $B_1$ is multiplied by $\alpha_1$ or $\alpha_2$,  which are both very small, any error will become negligible. We have in fact checked that this is so numerically. With this approximation we find

\be
- B_1(0,0,m_\phi) &=& -\frac{1}{2} \left(\ln \frac{Q^2}{m_i^2} -\ln \frac{m_{\phi}^2}{m_i^2} + \frac{1}{2} \right).
\ee

The Higgs contribution is the same for both the wino and bino. For simplicity, we take the mass of the lightest higgs boson to be zero in the $B$ functions (we do the same for the wino/bino again) and assume $m_H=m_{H+}=m_A$:

\be
\left(\frac{\Delta m_i}{m_i}\right)_{Higgs} &=& - \frac{\alpha_i}{4 \pi} \left[B_1(0,\mu,m_A)+B_1(0,\mu,0)+\frac{ \mu}{m_i} \sin{2 \beta}\left(B_0(0,\mu,m_A)-B_0(0,\mu,0)\right) \right]
\ee

\noindent Evaluating the $B$ functions, we find

\be
B_1(0,\mu,0) &=& \frac{1}{2} \left( \ln \frac{Q^2}{m_i^2} - \ln \frac{\mu^2}{m_i^2} + \frac{3}{2} \right)\\
B_1(0,\mu,m_A) &=& \frac{1}{2} \left[ \ln \frac{Q^2}{m_i^2} -\ln \frac {m_A^2}{m_i^2} + \frac{1}{2}+h(m_A^2/  \mu^2) \right]
\ee

\noindent   where $h(x) = \frac{1}{1-x}\left(1+\frac{\ln x}{1-x}\right)$. Note that $|h(x)|$ is a monotonically decreasing function of $x$. To estimate its maximum value, recall that $m_A^2 = 2 \bu / \sin 2 \beta$, so taking $\bu \sim \mu^2$ and setting $\tan \beta = 1$ implies $x = 2$, where $|h(x)| \sim 0.3$. This is already small, and its effect on the spectrum is negligible.  To keep consistent with the existing literature, we drop this term. Also,

\be
B_0(0,\mu,0) &=&  \ln \frac{Q^2}{m_i^2} - \ln \frac{\mu^2}{m_i^2} + 1\\
B_0(0,\mu,m_A) &=&  \ln \frac{Q^2}{m_i^2} -\ln \frac {m_A^2}{m_i^2} + \frac{\mu^2}{\mu^2-m_A^2} \ln \frac{m_A^2}{\mu^2} +1.
\ee

\subsection{NLO Formulae}
Adding in the two-loop beta function contribution to $m_3$, we arrive at the full NLO result

\be\label{eq:NLOg}
M_3 &=& m_3(Q) \left(1+\frac{3 \alpha_3}{4 \pi} \left[\ln \frac{Q^2}{m_3^2} + f(r)-\frac{14}{9} \right] +\frac{3 \alpha_t}{2 \pi} \right)\\
f(r) &=& 1+2r+2(r-1)^2 \ln |1-r| +2r(2-r) \ln r.
\ee

The NLO bino mass is

\be\label{eq:NLOb}
 M_1 &=& m_1(Q) \left(1+\frac{ \alpha_1}{8 \pi} \left[ -22 \ln \frac{Q^2}{m_1^2} + 11 \ln \frac{\mq^2}{m_1^2} + 9 \ln \frac{\ml^2}{m_1^2} +  \ln \frac{\mu^2}{m_1^2} +  \ln \frac{m_A^2}{m_1^2} \right. \right. \nonumber \\ 
 && \left. \left. {} +\frac{2 \mu}{m_1} \sin 2 \beta \frac{m_A^2}{\mu^2-m_A^2} \ln \frac{\mu^2}{m_A^2} -12  \right] +\frac{22 \alpha_3}{33 \pi}-\frac{13 \alpha_t}{66 \pi} \right) 
\ee
 
\noindent and the NLO wino mass is

 \be\label{eq:NLOw}
M_2 &=& m_2(Q) \left(1+\frac{ \alpha_2}{8 \pi} \left[ -2 \ln \frac{Q^2}{m_2^2} + 9 \ln \frac{\mq^2}{m_2^2} + 3 \ln \frac{\ml^2}{m_2^2} +  \ln \frac{\mu^2}{m_2^2} +  \ln \frac{m_A^2}{m_2^2} \right. \right.  \nonumber \\
 && \left. \left. {} +\frac{2 \mu}{m_2} \sin 2 \beta \frac{m_A^2}{\mu^2-m_A^2} \ln \frac{\mu^2}{m_A^2} +1.2 + 4.32 \ln \left(\frac{m_2}{m_W}-0.8 \right) \right] +\frac{6 \alpha_3}{ \pi}-\frac{3 \alpha_t}{2 \pi} \right).
\ee 

Notice that while our expression for the NLO gluino mass agrees with~\cite{gg}, there is a difference in the wino and bino masses, most notably in the coefficients of the $\ln Q^2$ terms. It seems that the authors of~\cite{gg} did not include a part of the higgsino/Higgs loop contribution in case of the bino, and omitted gauge boson loop contributions to the wino. For the sake of completeness, we have also included a two loop $\alpha_2$ contribution to the wino mass, which although it is small, should not in principle be dropped because it is of roughly the same size as the other terms.  We have confidence in our equations because they satisfy $\frac{d M}{d \ln Q}=0$ at one-loop order, after plugging in for the one-loop running of the gauge couplings. \\
\indent Finally, for the case of a wino LSP, we consider the splitting between the lightest chargino, i.e. $ \wt{W^\pm}$, and the LSP, $ \wt{W^0}$, which is important for understanding the phenomenology of the gluino cascade decay. Since the tree-level splitting due to mixing in the neutralino and chargino mass matrices is small for moderate to large $\mu$, the dominant contribution turns out to be due to gauge boson loops:

\be
\Delta m_{\wt{\chi} }=  m_{\wt{\chi}^+}-m_{\wt{\chi}^0}=\frac{\alpha_2}{2 \pi} \left[-2 B_0(m_2,m_2,m_W)+B_1(m_2,m_2,m_W)\right].
\ee

\noindent This is more conveniently expressed as 
\be
\Delta m_{\wt{\chi} } &=& \frac{\alpha_2 m_2}{4 \pi} \left[f(r_W)- \cos^2 \theta_W f(r_Z) - \sin^2 \theta_W f(0)\right],  \\
f(r_i) &=&  \int_{0}^{1} dx \, (2+2x) \ln [x^2+(1-x)r_i^2], \qquad r_i = \frac{m_i}{m_2}.
\ee
The splitting is roughly independent of $m_2$:  For $m_2 = 260$ GeV, $\Delta m_{\wt{\chi} } = 167$ MeV, while for $m_2 = 2.6$ TeV, $\Delta m_{\wt{\chi} } = 172$ MeV.

\section{Messengers and Sleptons}

To solve the tachyonic slepton problem, we introduce vector-like messenger fields in complete representations of $\mathbf{5} \oplus \mathbf{\bar{5}}$, which get masses from the following tree-level Kahler potential terms:

\be\label{eq:mess}
\lambda \int d^4 \theta \frac{ \phi^{\dagger}}{\phi} \mathbf{5  \bar{5}} +\kappa \int d^4 \theta \frac {X^{\dagger}X (\mathbf{5^{\dagger}5}+\mathbf{\bar{5}^{\dagger}\bar{5})}}{M_*^2}
 \ee
 
\noindent where $\phi$ is the conformal compensator, $X$ is the hidden sector field that breaks SUSY, and $M_*$ is a UV cutoff that is naturally the Planck scale in our model. In general we consider $N$ such sets of messenger fields, where we need $ N \leq 4 $ to preserve gauge coupling unification. Unification also works with one set of $\mathbf{5} \oplus \mathbf{\bar{5}}$ and one set of $\mathbf{10} \oplus \mathbf{\bar{10}}$ (we retain gauge coupling perturbativity in this case because the messengers have masses above 5 TeV~\cite{mp}). In fact, this is what we need for our complete model with the $\mu$ problem solution, to be described below. Since a $\mathbf{10}$ has a Dynkin index of $3/2$, this gives the same contribution to soft masses as a model with $N = 4$, as is clear from the formulae below.  \\
\indent Since it would be overly contrived to now introduce another scale in addition to $\mt$, we give the messengers masses of this order in a simple way through the EAM approach outlined in~\cite{nw}, which uses the Giudice-Masiero term~\cite{gm}. When supersymmetry is broken, the compensator acquires an $F$ term VEV so that $\phi = 1+m_{3/2} \theta^2$. The messengers also get soft masses through the second term in Eq. (\ref{eq:mess}). Since $m_{3/2} \sim F_{X}/ M_{Pl}$, the soft masses are also set by $\mt$. We parametrize the soft mass in terms of the supersymmetric mass $M$ as 

\be
m_{soft}^2 = -c^2 M^2, 
\ee

\noindent where the reason for the minus sign will become clear in a moment. To prevent the breaking of $SU(3)$, we require $ c \leq \sqrt{1-1/ \lambda}$, with $\lambda \geq 1$.  For simplicity, we assume the same GM coupling for all generations of messengers and equal soft masses for each messenger pair. Generalizing these assumptions does not significantly change the picture. We therefore have a hybrid theory of gauge and anomaly mediation, with messenger scale $M = \lambda m_{3/2}$,  and $F = - \lambda \mt^2$. We will examine the effect of the messengers on the gaugino spectrum in the next section. \\
\indent Here we focus on the soft masses, which give rise to Poppitz-Trivedi D-type gauge mediation. This mechanism was used to solve the tachyonic slepton problem in~\cite{chacko,nws,ga}. The idea is simple: since the scalars and messengers share gauge interactions, soft masses for the messengers will induce scalar masses. This contribution was calculated by Poppitz and Trivedi~\cite{pop} to be 

\be  
\Delta m_i^2 = - \sum_{a} \frac{g_a^4}{128 \pi^4} S_M \, C_{ai} \, Str M_{mess}^2 \, \mathrm{log} \frac{\Lambda^2}{m_{IR}^2},
\ee

\noindent where $S_M$ is the Dynkin index of the messenger field, $C_{ai}$ is the quadratic Casimir, and $\Lambda, m_{IR}$ are UV and infrared cutoffs, respectively. The logarithm is large, since we take $\Lambda$ at the GUT scale and $m_{IR} = M$ (The natural cutoff for our model is the Planck scale, so this is not the entire contribution. There is also a correction coming from physics between the GUT and Planck scales that is not log-enhanced, and includes unknown threshold corrections at the GUT scale). As pointed out in~\cite{nws}, the large logs can be resummed using the one-loop RGE's for the gauge couplings, yielding

\be
\Delta m_i^2 = - \sum_{a} \frac{S_M \, C_{ai} \, Str M_{mess}^2[g_a^2(\Lambda)-g_a^2(m_{IR})]}{8 \pi^2 b_a},
\ee
 
\noindent with $b_a$ the $\beta$ function coefficient above the messenger scale. In our case, $Str M_{mess}^2 = 4 m_{soft}^2$, so we need $m_{soft}^2$ to be negative in order to get a positive contribution, hence the minus sign in the definition above. Since $\Delta m_i^2 \sim (\Delta g^2/16 \pi^2) (c \lambda)^2 \mt^2 N$, it is clear that this contribution can easily be as large as that from AMSB for relatively small values of $c \lambda$, i.e. they need not even be $O(1)$, pushing the slepton masses positive at the messenger scale. 

\section{Gaugino Spectrum}
We now take a closer look at how the messengers change the gaugino spectrum. Integrating out the messengers takes the soft terms off of the anomaly mediated trajectory. This means that to get the gaugino masses at the weak scale, we need to compute them first at the messenger scale and then run down. We first run up the gauge couplings at two loops to do this. Immediately above the messenger scale the gaugino masses are on the anomaly-mediated trajectory. Using the two-loop beta functions, which include contributions from the messengers, we find  

\be
m_1(M) &=&  \frac{\alpha_1}{4\pi} \left(11+\frac{5}{3}N + \frac{\alpha_2}{4\pi} (3N+9) + \frac{\alpha_3}{4\pi} \left(\frac{32}{9}N+\frac{88}{3} \right)-\frac{13 \alpha_t}{6\pi}\right) m_{3/2} \\
m_2(M) &=&  \frac{\alpha_2}{4\pi} \left(1+ N + \frac{\alpha_2}{4\pi} (7N+25) + \frac{24 \alpha_3}{4\pi}-\frac{3 \alpha_t}{2\pi}\right) m_{3/2} \\ 
m_3(M) &=&  \frac{\alpha_3}{4\pi} \left(-3 + N + \frac{ 9 \alpha_2}{4\pi}  + \frac{\alpha_3}{4\pi}\left(\frac{34}{3}N + 14 \right) - \frac{\alpha_t}{\pi} \right) m_{3/2}.
\ee
 
\noindent Since $\alpha_3 \sim 3 \alpha_2$ at the weak scale, we see that we can make the gluino and wino masses approximately equal at the messenger scale by choosing $N = 2$. There are also threshold corrections from integrating out the messengers, and the exact expression depends on whether or not the messengers have soft masses. Here we first consider the more general case of messenger soft masses, which are needed in our model. We then discuss the simpler case with no soft masses.
\subsection{Soft Masses}
Adapting the formula in~\cite{pop}, we find

\be
\Delta m_i = \frac{\alpha_i}{2 \pi} f_t(y_1,y_2) N \mt, \qquad f_t(y_1,y_2) = \frac{y_1 \mathrm{log} y_1 - y_2 \mathrm{log} y_2 - y_1 y_2  \mathrm{log}(y_1/y_2)}{(y_1 - 1) (y_2 - 1) (y_2 - y_1)}
\ee 
 
\noindent with $y_1 = M_1^2/M^2$, $y_2 = M_2^2/M^2$, where $M_{1,2}^2$ are the eigenvalues of the scalar messenger mass-squared matrix (we adopt a convention where $M_1$ is the larger of the two). For the simplest case of universal GM couplings and soft masses that we consider

\be
y_1= 1- c^2+1/\lambda, \qquad y_2= 1- c^2-1/\lambda.
\ee

For $\lambda$ not too close to 1 and small soft masses, $f_t \sim 0.5$. As $\lambda \to 1$ (soft masses still small), $f_t \to 0.7$ since there is an enhancement due to contributions from higher order terms in $F/M^2$~\cite{martin}. To examine the role of the soft masses, it is useful to consider the effective number of messengers $N_{eff}$, a continuous variable defined by

\be
N_{eff} = \frac{1+2 f_t}{2} N.
\ee

\noindent Here we consider both positive and negative $m_{soft}^2$ for $\lambda$ not too close to 1. In the positive case, as can be seen in Fig. 1, $N_{eff}$ decreases with increasing $c$.  As $c$ becomes bigger the soft mass dominates over the supersymmetric mass and $N_{eff} \to N/2$. Note that with positive $m_{soft}^2$, $c$ is not constrained to be smaller than one. $N_{eff}$ increases with $c$ for negative $m_{soft}^2$, although it is very gradual until $c \sim 0.6$, as shown in Fig. 2.

\begin{figure}
\centering
\includegraphics{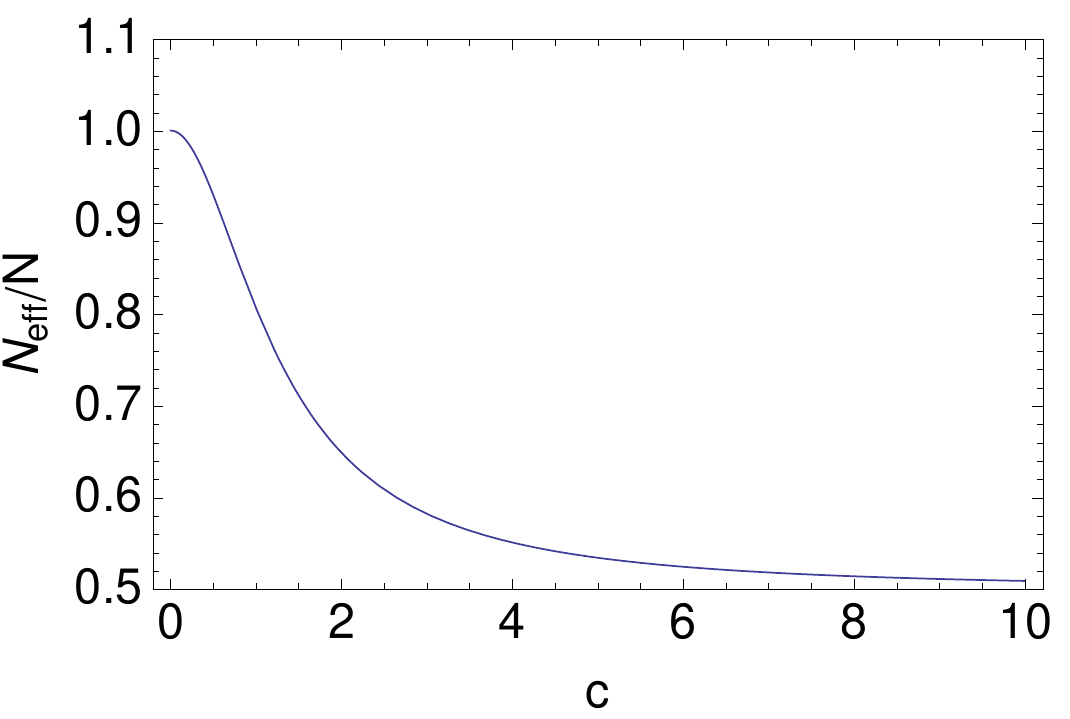}
\caption{Effect of c for positive $m_{soft}^2$}
\end{figure}

\begin{figure}
\centering
\includegraphics{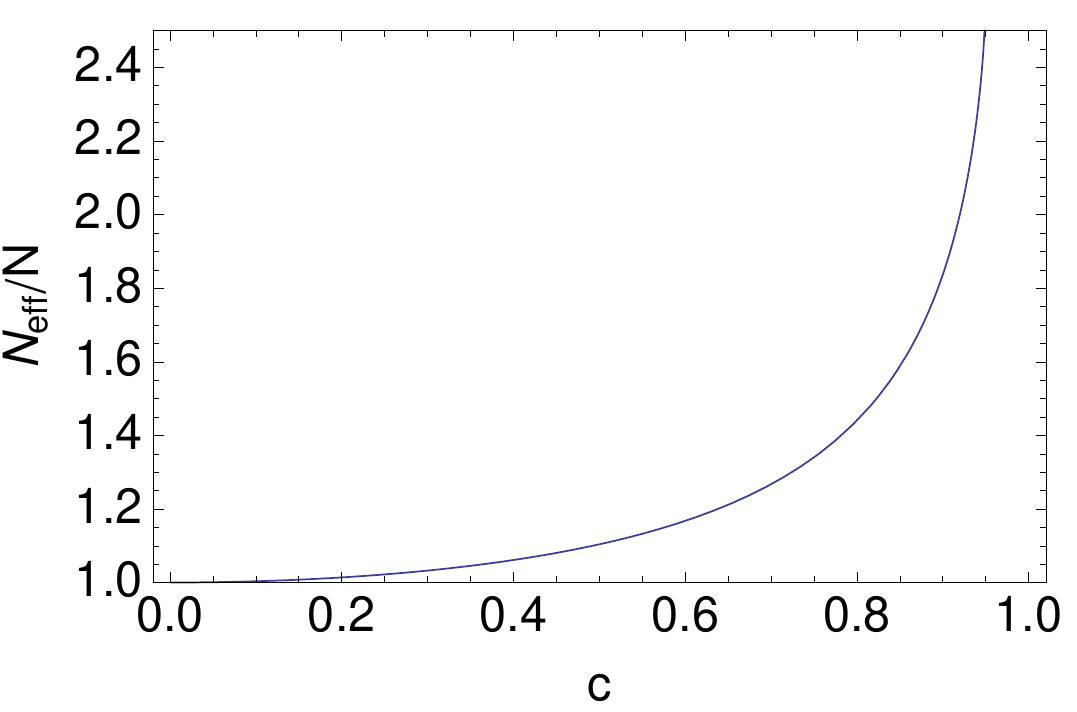}
\caption{Effect of c for negative $m_{soft}^2$}
\end{figure}

\subsection{No Soft Masses}
The compression of the gaugino spectrum due to messengers is an interesting aspect of extended AMSB theories that, according to our knowledge, has not been investigated previously. We therefore consider it now as an independent module that can be incorporated into other models, i.e. we just focus on the gaugino masses. In this case there is no reason to keep the messenger soft masses, so we simplify the setup by just keeping the GM term for the messengers. In this case, the threshold correction from integrating out the messengers takes the simpler form~\cite{luty2}

\be
\Delta m_i (M )= \frac{1}{g_i}(\beta_{i}'-\beta_i) G(F/M^2) \frac{F}{M},
\ee

\noindent where $ \beta'$ is the beta function above the messenger threshold, $ \beta$ is the beta function below the messenger threshold, and $G$ is the enhancement factor mentioned above.  G($x$) increases monotonically from 1 to 1.386 as $x$ goes from 0 to 1~\cite{martin}. Here we also take different couplings $\lambda_D$ and  $\lambda_T$  for the triplets and doublets, although the coupling cancels out in $F/M$, which means that we can only adjust the value of the higher order contribution $G$ separately for the triplets and doublets. Explicitly, the threshold corrections to two-loop order are

\be
\Delta m_1 &=&  \frac{\alpha_1}{4\pi} \left(\frac{5}{3} + \frac{3 \alpha_2}{4\pi}  + \frac{8 \alpha_3}{9\pi}\right)  \left[G(x_D)+ \frac{2}{3} G(x_T) \right] \frac{3}{5} N m_{3/2} \\
\Delta m_2 &=&  \frac{\alpha_2}{4\pi} \left(1 + \frac{7 \alpha_2}{4\pi}\right) G(x_D) N  m_{3/2} \\
\Delta m_3 &=&  \frac{\alpha_3}{4\pi} \left(1 + \frac{17 \alpha_3}{6\pi}\right) G(x_T) N m_{3/2},
\ee
 
\noindent where $x = -1/ \lambda$. \\
\indent We then run down the gaugino masses to 1 TeV using the one-loop RGEs (we include the next-to-leading order correction for the gluino) and compute the pole masses by adding the corrections appearing in the square brackets in Eqs. (\ref{eq:NLOg}), (\ref{eq:NLOb}), and (\ref{eq:NLOw}). To keep the log term from getting too large, we compute the gluino pole mass at a separate scale, equal to the running mass at $1$ TeV. Here we use the $\overline{MS}$ equations so the pole mass equation for the gluino is modified to 

\be
M_3 &=& m_3(Q) \left(1+\frac{3 \alpha_3}{4 \pi} \left[ \ln \frac{Q^2}{m_3^2} + f(r) - 1 \right] \right).
\ee

\begin{table}[t]\label{table:gaugino}
\begin{center}
\begin{tabular}{|c|c|c|c|c|c|c|c|c|c|c|}
\hline
N & $\mt$ (TeV) & $\ld$ & $\lt$ & $M_1$ (GeV) & $M_2$ (GeV) & $M_3$ (GeV) & $\Delta$M (GeV) \\
\hline
1 & 70 & 2.5 & 2.5 & 851 & 618 & 523 & $\tilde{g}$ lightest  \\
2 & 40 & 2.5 & 1.2 & 608 & 580 & 650 & 74  \\
3 & 40 & 1.5 & 4.0 & 721 & 818 & 1130 & 411  \\
4 & 30 & 1.1 & 4.0 & 643 & 831 & 1309 & 668  \\ 
\hline
\end{tabular}
\caption{Gaugino spectra}
\end{center}
\end{table}

Finally, to calculate the splitting between the gluino and the LSP, we diagonalize the neutralino mass matrix using the calculated pole masses for the bino and wino.

We now discuss the gaugino spectra for different numbers of messengers. Examples are presented in Table 1, with $M_3$ the gluino mass and $\Delta{M}$ the gluino-LSP splitting. We assume that the scalars and the higgsinos are somewhat heavier than the gluino; here we choose an arbitrary mass of 1.5 TeV for all, and take tan $\beta = 5$. With heavy squarks, the dominant SUSY production mechanism at the LHC is gluino pair production. Each gluino then eventually decays to the bino or wino LSP. In either case, direct decay to the LSP, $\wt{g} \to j j \wt {\chi}^0$, is possible and is the dominant mode for a bino LSP. For the case of a wino LSP, there is also cascade decay through the charged wino, $\wt{g} \to j j \wt {\chi}^\pm$. For $\Delta m_{\wt{\chi} } > m_\pi$, $ \wt{\chi}^\pm \to \pi^{\pm} \wt{\chi}^0$ happens $98 \%$ of the time. For the range of wino masses that we consider $\Delta m_{\wt{\chi} }$ is roughly 170 MeV, so this mode is always open. Although the charged wino can travel a macroscopic distance before decaying~\cite{gg} (about 1 cm in our case), a displaced vertex analysis is not possible because the pion is too soft. Thus, in terms of observable signatures, we can simply describe the decay to a wino LSP as also being direct. We refer mainly to~\cite{cms12,at12_13} to determine which parts of parameter space are still available for such a simplified direct decay model for moderate compression, and to~\cite{atf} in the extremely compressed case.  \\
\indent As pointed out in~\cite{wacker}, despite the large production cross-section, these events are difficult to detect when the gluino is nearly degenerate with the LSP since the jets from the decay are very soft. Furthermore, these events may not even have the large $E_{T}^{miss}$ that is usually a hallmark of R-parity conserving theories, meaning that they will be hidden in QCD background. Even if the gluinos are strongly boosted the LSP momenta will approximately cancel unless the gluino momenta are unbalanced by the emission of initial or final state radiation. \\ 
\indent For $N = 1$, we do not obtain an acceptable spectrum--the gluino is always the lightest. This is because in this case the contributions to the gluino mass from anomaly mediation and the messenger threshold correction have opposite sign. Although the correction due to squarks is sizable, bumping up the gluino substantially would require very heavy squarks. This is because the squark correction increases slowly as the ratio of the squark mass to $m_3$ is raised, as can be seen in Fig. 3. 

\begin{figure}
\centering
\includegraphics{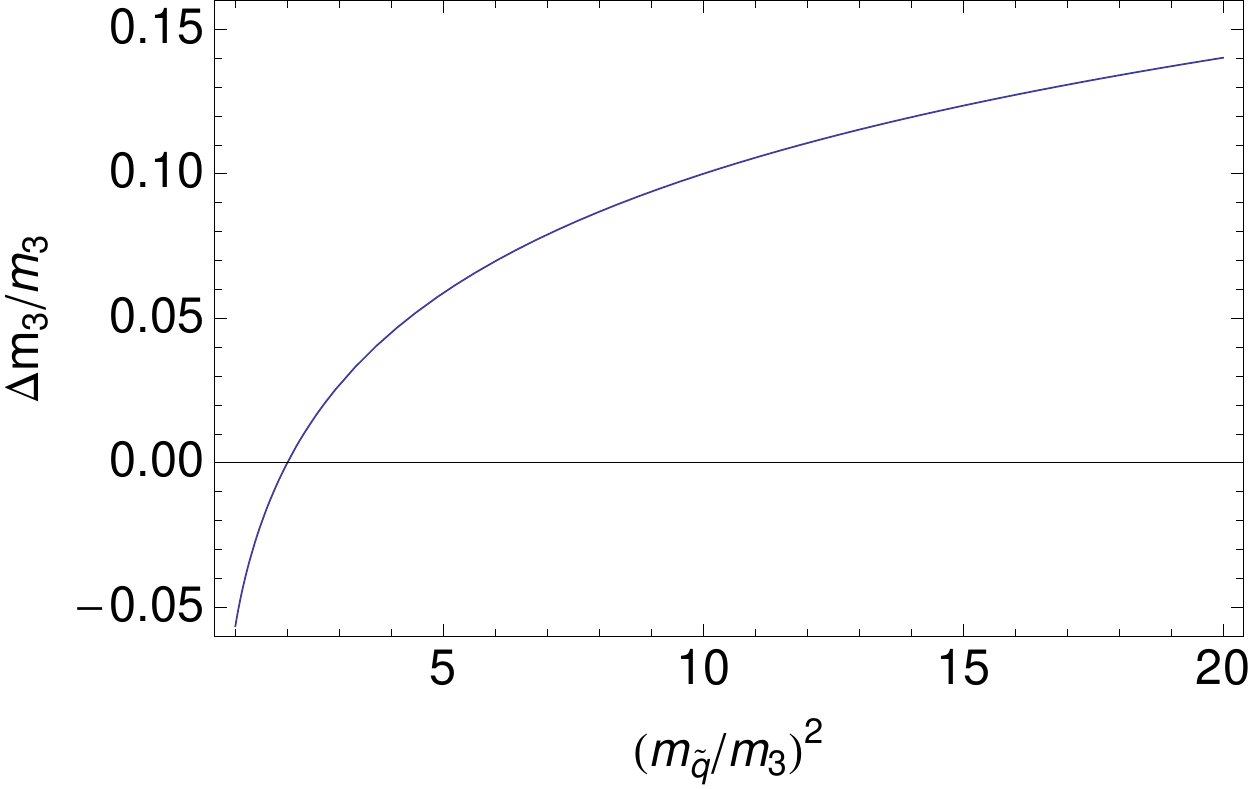}
\caption{Squark contribution to gluino mass}
\end{figure}
\indent For $N = 2$, the wino is the LSP unless $\lambda_D$ gets very close to 1, in which case it is the bino. The spectrum is very compressed, with $\Delta M$ no bigger than about 80 GeV. As can be seen in Fig. 4, $\dm$ increases with $\ld$ because the messenger threshold correction to the wino gets smaller, decreasing its mass. Conversely, decreasing $\ld$ raises the wino mass more than the bino mass because of the smallness of $\alpha_1$, eventually pushing the wino above the bino for $\ld ~\sim1$. 

 \begin{figure}
\centering
\includegraphics{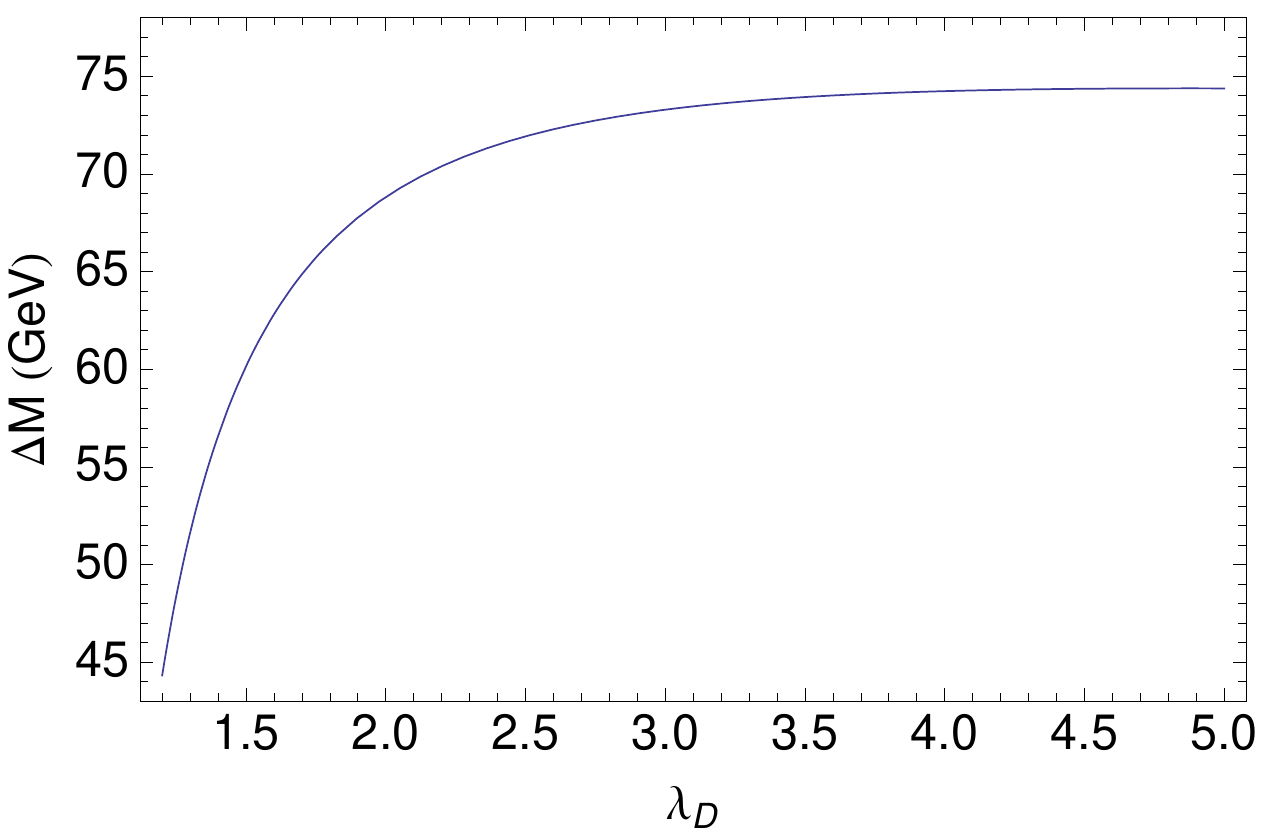}
\caption{Gluino-LSP Splitting for N = 2}
\end{figure}

\indent The $N = 3$ and $N = 4$ cases are not as compressed but are still squeezed substantially compared to pure AMSB. The bino is the LSP in both cases. We focus on gluino masses below 1.4 TeV, since this is the region that many claim has already been ruled out. Varying $\ld$ has only a slight effect because $\dm$ is a few hundred GeV whereas $ \Delta m_1 $ is about $200$ GeV, meaning that $\dm$ can only change by a few tens of GeV.  \\
\indent In summary, for 2, 3, or 4 messenger pairs, relatively light gluinos in the range of $600$ to $1300$ GeV are still viable.  Additionally, compression of the gaugino spectrum is attractive in Split Supersymmetry, where the gluino is perhaps the only sign of new physics and is out of the reach of the LHC with the usual AMSB mass hierarchy and wino/bino thermal dark matter. We will discuss incorporating messengers into the Split SUSY framework in a forthcoming publication~\cite{tz}.\\

\section{Complete Example Spectra}

In the previous sections we have discussed separately the solution to the tachyonic slepton problem through D-type gauge mediation and the compression of the gaugino spectrum from the messenger threshold in EAM. We now combine the two to produce complete spectra for different numbers of messengers in the theory. These are listed in Table 2. We assume that a solution to the $\mu$ problem exists (more on this in the next section) and take $\mu = 1$ TeV (with $B \mu \sim \mu^2$) and three values of $\tan \beta$. For these chosen parameters, we calculate the additional contributions to the higgs soft masses that are needed for proper EWSB and add them at the messenger scale. 

\begin{table}[t]\label{table:points}
\begin{center}
\begin{tabular}{c|c|c|c|c|}
& & $N = 2$ & $N = 3$ & $N = 4$  \\
\hline
inputs:  &
$\mt$ & 40000 & 40000 & 30000 \\
&$\lambda$& 1.3 & 2.5 & 3.0  \\ 
&$c$& 0.25 & 0.10 & 0.09  \\ 
&$\tan \beta $& 10.0 & 13.0 & 3.0  \\ 
&$\mu$& 986 & 984 & 979 \\
\hline
sleptons:
&$m_{\tilde{e}_L}$  &750 & 762 & 944   \\
&$m_{\tilde{e}_R}$  &657 & 754 & 650     \\
&$m_{\tilde{\nu}_L}$ &750 & 762 & 944  \\
\hline
squarks:
&$m_{\tilde{u}_L}$ & 1880 & 1959 & 2186   \\
&$m_{\tilde{u}_R}$ & 1704 & 1752 & 2016 \\
&$m_{\tilde{d}_L}$ & 1880 & 1959 & 2186\\
&$m_{\tilde{d}_R}$ & 1732 & 1805 & 1998  \\
\hline
stops:
&$m_{\tilde{t}_1}$ & 1532 & 1565 & 1827  \\
&$m_{\tilde{t}_2}$ & 1816 & 1889 & 2112   \\
\hline
gauginos:
&$m_{\tilde{B}}$ & 619 & 714 & 623 \\
&$m_{\tilde{W}}$ & 616 & 795 & 769  \\
&$m_{\tilde{g}}$ & 703 & 1193 & 1396  \\
\hline
Higgs sector:
&$m_A$ & 5727 & 5715 & 3209 \\
\end{tabular}
\caption{Example spectra for $N = 2,3,4$. All masses are in GeV.}
\label{tab:points}
\end{center}
\end{table}

In the $N = 2$ and $N = 3$ cases, we choose squarks in the range of $1.7 - 1.9$ TeV, with a slightly lighter, right-handed stop (there is little mixing), and sleptons that are lighter by about a factor of two. The N = 4 case has slightly heavier squarks compared to these. In principle, the squarks could be much lighter in these scenarios, and sneutrinos would become the LSP (a spectrum we do not study, as we are investigating scenarios with lighter gauginos). Since the D-type GMSB contribution dominates over its AMSB counterpart -- it is about an order of magnitude larger -- the mass hierarchy can be explained by ratios of gauge couplings, as in minimal GMSB. However, the near equality of the slepton masses in the $N = 2$ and $N = 3$ cases is a departure from this and arises from both EWSB constraints and our choice of larger $\tan\beta$.  Because $\mu$ is fairly big, we need $\mhd$ to be large to obtain sizable values of $\tan\beta$. In addition, $\mhu$ must be negative to satisfy the other EWSB equation involving $m_Z$. This means that the oft-neglected $\alpha^2_1(\mhu-\mhd + ...)$ piece in the RGE for $m^2_{\tilde{e}_R}$ is significant and drives the mass up considerably when running down. In the $N = 4$ case, we chose a smaller $\tan\beta$ so $\mhd$ is not as large and $\mhu$ is positive, and this effect is greatly diminished. 

The gaugino masses are similar to those in Table 1 because the messenger soft masses have only a slight effect on the messenger threshold correction for the small values of $c$ that we need.  As noted before, we ignore the $N = 1$ case since that results in a gluino LSP. The mechanism that generates $\mu / B \mu$ will obviously have an impact on the physical higgs mass, so we do not worry about that here. 
 

\section{Solving the $\mu$ Problem}

It is well known that one cannot write down a tree-level $\mu$ term in minimal AMSB because the resulting $B \mu$ would be a loop factor too large to allow for proper EWSB. However, D-type gauge mediation is an extra element in our model that contributes to the Higgs soft masses. Taking inspiration from~\cite{nom}, we examine whether we can use this extra freedom to increase the Higgs soft masses enough to make EWSB work despite the large $\mu/ \bu$ hierarchy. We find roughly that to barely satisfy the EWSB stability condition (tan $\beta = 1$), $ c \lambda \sim 3 $, yielding squark masses close to 15 TeV, which is clearly not acceptable.   Since the two Higgs doublets can be considered a ``messenger pair", one can also try to generate $\mu$ and $\bu$ by writing down a GM term for them, as originally proposed. However, $\bu$ again generally turns out to be a loop factor larger than $\mu^2$, so we face the same problem. It is clear that we must add something new to our model. \\
\indent We can try to solve the $\mu$ problem by taking the simplest extension, one that was considered in the early days of gauge mediation~\cite{pom}. We generate $\mu/ \bu$ by coupling the messengers to the Higgs doublets in the following way:

\be
W \supset z_u \mathbf{\bar{10}} \, \mathbf{5} H_u  + z_d \mathbf{10} \, \mathbf{\bar{5}} H_d.
\ee

\noindent This again leads to the same $\mu/ \bu$ hierarchy, but also gives an extra contribution to the Higgs soft masses, so that we don't have to rely solely on the GMSB contribution. Working to all orders in $F/M^2$ and including the soft masses, the new yukawa couplings generate the following contributions to the Higgs sector:

\be
\mu &=& \frac{z_u z_d}{2 \pi^2} f_t(y_1,y_2) \mt \\
\bu &=&  -\frac{z_u z_d}{4 \pi^2} \ln(1-x^2) \lambda^2 \mt^2  \\
\Delta m_{h_{u,d}}^2 &=&  -\frac{z_{u,d}^2}{4 \pi^2} \left[\left(2-c^2 \right)\ln(1-x^2) +\left(1-c^2 \right) x \, \ln \frac{1+x}{1-x} \right] \lambda^2 \mt^2 ,  \\
x &=& \frac{1}{\lambda(1-c^2)} \, .
\ee

\noindent For illustrative purposes only, we also give the formulas for the soft parameters to lowest order in $1/\lambda^2$ and $c^2$. These should roughly show the correct qualitative behavior since $ c < 1 $ and $ \lambda \geq 1$:

\be
\bu &=&  \frac{z_u z_d}{4 \pi^2}(1+2 c^2) \mt^2  \\
\Delta m_{h_{u,d}}^2 &=& \frac{z_{u,d}^2}{4 \pi^2} c^2  \mt^2.
\ee

\noindent As first noted in~\cite{pom}, it is clear from the above that the Higgs soft masses do not get a contribution at lowest order with messenger soft masses equal to zero. Because of this fact, we must introduce a hierarchy between $\zu$ and $\zd$ so that $\tan \beta$ is not fixed too closely to one, i.e. we need $m_{h}^2  > \bu$ for one of the higgs doublets. So the EWSB will be achieved using the approach outlined in~\cite{nom}, with a large $\mu / \bu$ hierarchy and a smaller one between $\bu$ and either $m_{h_{u}}^2$ or $m_{h_{d}}^2$. Taking this to be $h_u$, we find that $\sin 2\beta \sim 2(z_d/z_u)$. Since we do not want higgsinos that are too light, we must fix the product $ \zu \zd $. However, we cannot make $z_u$ larger than 1 without hitting a Landau pole before the GUT scale. These constraints imply that $\mu$ is small, about 300 GeV. The gluino mass with a scalar spectrum similar to that of the $N = 2$ or $N = 3$ cases in Table 2 is then about 1.5 TeV. Although this spectrum is viable (barely), it is only because of the size of the gluino mass and not because of compression, so we do not find this attempt at a $\mu$ problem solution to be satisfactory in this minimal form. Note also that we need to tune the value of the other yukawa coupling so that the condition for EWSB in this case, $ \bu^2 > m_{h_{u}}^2 m_{h_{d}}^2$, nearly becomes an equality. This has to be done so that we obtain the experimental value of $m_Z$ in the other EWSB equation. \\
\indent A further issue is the physical higgs mass. In our model with all messenger soft masses negative, the scalars in each messenger multiplet are lighter than the fermions, so the new higgs couplings will produce a negative contribution to the higgs mass squared, whereas to achieve a higgs mass of 125 GeV~\cite{higgs} we need a large, positive correction for our case of $\sim 2$ TeV stops~\cite{meade}. It seems this can be overcome by choosing different soft masses for the $\mathbf{5}$ and $\mathbf{10}$ multiplets, with the scalar soft masses of the $\mathbf{10}$ negative so that the D-type GMSB contribution to the MSSM scalar masses is still positive. Analogously to the top/stop contribution, the messenger contribution to the higgs mass depends on the logarithm of the ratio of the geometric average of the scalar masses to the fermion mass. So if the  scalars from the $\mathbf{5}$'s have positive soft masses that are larger in magnitude than those of the scalars from the $\mathbf{10}$'s, the average messenger scalar mass can be made larger than the fermion mass, yielding a positive contribution to the higgs mass.\\
\indent To improve our solution, we clearly need an additional contribution to $\mu$. This can be done by introducing a singlet and working in the context of the NMSSM. This has been investigated recently in the context of GMSB with new messenger couplings in~\cite{comp}. In this case the physical higgs mass results from the large A terms that are produced in this model. Since the singlet is not charged under the SM gauge groups, it does not affect the AMSB or GMSB contributions to the soft masses. In particular, the gaugino spectrum remains unchanged. This model could then eventually be realized in a 5D brane-world setup similar to that of~\cite{chacko,nws}, where the visible (MSSM) and hidden sectors are localized on different branes, with the gauge fields and the messengers in the bulk. 

\indent 
\indent 
\indent 


\section{Conclusions}
The most recent data from the LHC excludes gluinos with masses less than $\sim 1.4$ TeV in typical models that have a significant gluino-LSP mass splitting, putting a strain on naturalness. However, gluinos as light as 550 GeV are still allowed for very small mass splittings. We have presented a simple and novel way that such a compressed gaugino spectrum occurs naturally in the context of AMSB. \\
\indent AMSB models typically have spectra that feature a large spread in the gaugino masses $\--$ the gluino is almost 10 times heavier than the wino LSP. Such models can be out of the reach of the LHC for LSP masses of $\cal{O}$(1 TeV). We find that the presence of messengers in the AMSB framework compresses the spectrum. The resulting spectra have a relatively light gluino with a mass in the range of 600 to 1300 GeV that is no heavier than about twice the LSP mass, with the exact values dependent on the number of messengers $N$ used. For the $N=1$ case, the gluino is the LSP,  while the $N=2$ case yields a wino or bino LSP depending on the value of the coupling in the Giudice-Masiero term for the messenger doublets, and the mass splitting between the gluino and the LSP is of the order of tens of GeV. The $N=3,4$ cases are less compressed and yield a bino LSP. \\
\indent We have provided expressions for the gaugino pole masses which differ from the expressions present in the literature for the case of the wino and bino. We would like to emphasize that we have confidence in our expressions being correct as the pole masses are independent of the running scale in our case. We have discussed in detail the steps to obtain the pole masses to account for the differences. \\
\indent Apart from compressing the spectrum, the messengers are crucial in building a complete phenomenological model without singlets, as they help solve the tachyonic slepton problem in AMSB in a way previously suggested in the literature. Contact terms between the messenger fields and hidden sector SUSY breaking result in soft masses for the messengers. Through Poppitz-Trivedi D-type gauge mediation, these soft masses generate contributions to scalar masses which are positive if the soft masses squared are taken to be negative. This contribution is of the same order or greater than the AMSB contribution and thus solves the tachyonic slepton problem. The only hurdle to a complete model that then remains is the $\mu$ problem. We have presented a simple solution as an example that proves that solutions do exist in our framework. It involves new yukawa couplings between the Higgs doublets and messengers (i.e. not new fields). However, the minimal version that we considered is deficient, since it produces a light higgsino LSP and requires considerable fine-tuning. We leave open for future work possible extensions with both positive and negative messenger soft masses and/or the NMSSM as an avenue for resolving the remaining problems.

\indent 
\indent 

\section{Acknowledgements}

A.G. and T.Z. would like to thank Fabrizio Caola, Liang Dai, Andrei Gritsan, Gordan Krnjaic, Jingsheng Li, Kirill Melnikov and Andrew Whitbeck for helpful discussions.  This work was supported in part by the National Science Foundation, grant number PHY-0244990.  




\end{document}